\documentclass[prb,twocolumn,showpacs,citeautoscript]{revtex4}

\usepackage{graphicx}% Include figure files
\usepackage{dcolumn}% Align table columns on decimal point
\usepackage{bm}% bold math
\usepackage{ulem}%pretty font effects

\begin{document}

\preprint{APS/123-QED}

\title{Ideal barriers to polarization reversal and domain-wall
motion in strained ferroelectric thin films}

\author{S. P. Beckman}
\email{sbeckman@iastate.edu}
\thanks{Present address:Department of Materials Science and
Engineering, Iowa State University, Ames, IA 50011-2300.}
\author{Xinjie Wang}
\author{Karin M. Rabe}
\author{David Vanderbilt}

\affiliation{Department of Physics and Astronomy, Rutgers
University, Piscataway, New Jersey 08854-8019}

\date{\today}% It is always \today, today,

\begin{abstract}
The ideal intrinsic barriers to domain switching in $c$-phase
PbTiO$_3$ (PTO), PbZrO$_3$ (PZO), and PbZr$_{1-x}$Ti$_x$O$_3$
(PZT) are investigated via first-principles computational methods.
The effects of epitaxial strain on the atomic structure,
ferroelectric response, barrier to coherent domain reversal,
domain-wall energy, and barrier to domain-wall translation are
studied.  
It is found that PTO has a larger polarization, but smaller energy 
barrier to domain reversal, than PZO.  
Consequentially the idealized coercive field is over two times smaller 
in PTO than PZO.  
The Ti--O bond length is more sensitive to strain than the other bonds 
in the crystals.  
This results in the polarization and domain-wall energy in PTO having 
greater sensitivity to strain than in PZO.  
Two ordered phases of PZT are considered, the rock-salt structure 
and a (100) PTO/PZO superlattice.  In these simple structures we
find that the ferroelectric
properties do not obey Vergard's law, but instead can be approximated 
as an average over individual 5-atom unit cells. 
\end{abstract}

%Use showkeys class option if keyword

\maketitle

\section{Introduction}

The advent of advanced techniques for the growth of high-quality
ferroelectric thin-films \cite{posadas2007}
has spawned a
great deal of interest in using these films for
high-density non-volatile ferroelectric information storage
\cite{Dawber2005,Scott2007,Arimoto2004}.
The device concept involves
associating 0 and 1 bits with ``up'' and ``down'' polarized
domains in the film\cite{Kato2005,Arimoto2004}.

For this application, assuring the retention of the data is of critical
importance\cite{Arimoto2004}. It is necessary therefore
to establish an understanding of
the factors that govern the stability of polarization domains.
As will be seen from the discussion below, the domain-wall energy
and mobility are especially important. One
promising avenue to control
these critical factors is by tuning the epitaxial strain in the
thin films, present
due to lattice mismatch between the film material
and the substrate.
Both the investigation of the stability mechanisms
and of the effects of epitaxial strain can be carried out
using first principles methods, as will be described further below.

Polarization switching is believed to occur via the nucleation of
an oppositely polarized domain at a defect such as a dislocation
or inclusion, followed by domain wall advancement outward from
the embryonic domain.  The wall does not propagate uniformly,
but by a mechanism of kink nucleation and motion initiated at
a defect or compositional inhomogeneity where the local barrier
is reduced.  Subsequently this small region
expands lateral to the original wall, converting the phase as it
passes\cite{Shin2007}.  This process is complicated and depends
strongly on extrinsic effects such as the quality of the film,
the processing conditions, the mechanical, thermal, and electrical
histories, as well as the environmental conditions at the surface
of the film.

Underpinning the extrinsic behavior is the intrinsic, atomic
nature of the ferroelectric material.  We focus here on two
levels of idealized intrinsic behavior.  First, we consider the
barrier for coherent domain reversal, in which every unit cell in the
crystal undergoes simultaneous reversal from one ferroelectric
ground state to its opposite one.  This is a purely bulk property,
since the crystal is assumed to maintain its three-dimensional
periodicity throughout, and the barrier is determined by a saddle
point corresponding to the centrosymmetric (paraelectric) bulk
structure.

Second, we consider the idealized barrier for a domain
wall to translate by one unit cell in the direction normal to the
wall, while always maintaining two-dimensional periodicity in
the parallel directions.  This ideal barrier for domain-wall
motion is determined by a saddle-point structure in which the
domain wall is centered at an unfavorable position in the unit
cell.  In both cases, we focus on zero-temperature transition
paths. First-principles calculations have
been performed to examine 180$^{\circ}$ domain walls in PbTiO$_{3}$
\cite{Meyer2002}, and valence-bond methods have been employed
to study domain-wall motion in PbZr$_{1-x}$Ti$_{x}$O$_{3}$
\cite{Shin2007}.

The energy to reverse the polarization via the coherent-reversal
or domain-wall-translation path is proportional
to the volume of the film or to the area of the domain wall respectively,
and thus may be orders of magnitude larger than the true thermal
barrier for motion of the domain wall.  However, even if it is
not quantitatively relevant to the
real mechanisms of domain-wall motion, the study of the
idealized barriers give insight into the underlying atomistic
mechanisms and materials properties that
are relevant to real domain-wall motion.  Thus, any more sophisticated
treatments that may be put forward in the future should start
from a firm understanding of these intrinsic
effects.

The impact of strain on a crystal's ferroelectric response
has been theoretically studied using both phenomenological
and first principles techniques.  For single-phase
compounds, epitaxial strain phase diagrams have been obtained both 
from the Landau-Devonshire free-energy\cite{Pertsev1998}
and from first-principles
effective-Hamiltonian methods\cite{Dieguez2005,King-Smith1993}.
The effect of strain has also been studied in superlattices
\cite{Bungaro2004}.  However the effect of composition and
strain on polarization reversal and domain-wall energy has yet
to be reported in detail.

In this paper, the ideal barriers to coherent polarization reversal
and domain-wall motion are studied for $c$-phase PbTiO$_3$ (PTO),
PbZrO$_3$ (PZO), and two ordered forms of PbZr$_{1-x}$Ti$_x$O$_3$
(PZT), the rock-salt structure and a (100)-oriented
superlattice.
In particular, we report
the dependence of the structure, the spontaneous ferroelectric
polarization, the barrier to coherent domain reversal, the
domain-wall energy, and the barrier to domain-wall translation
as a function of epitaxial strain.

The manuscript is organized as follows.  In Sec.~\ref{TA} we
discuss the boundary conditions used in our approach,
present the geometries of the cells used to treat pure
PTO and PZO and the ordered PZT alloys, and give the
details of the computational methods that were used.
The results of the calculations are presented
in Sec.~\ref{results}.  A more detailed analysis of the crystal
structures and an identification of the relevant atomistic features is
given in Sec.~\ref{D}.  Finally, we summarize and conclude
in Sec.~\ref{C}.

\section{Theoretical Approach\label{TA}}

\subsection{Boundary conditions}

In the study of ferroelectric thin films, one has to
specify both the mechanical boundary conditions, such as epitaxial
strain constraints, \cite{Dieguez2005}
as well as the electrical boundary conditions.
In the present calculations, periodic boundary conditions
are used. This corresponds to zero macroscopic field such as would
be obtained with ideal symmetric short-circuited electrodes. Thus,
these results may not apply in film geometries
in which an electrical bias is intentionally or unintentionally
present\cite{Dawber2003,Junquera2003,Yoshida1999}.
We further note that these periodic boundary conditions 
produce ``constrained bulk" calculations, similar to those of 
Ref.~[\onlinecite{Dieguez2005}],
in which we carry out a bulk calculation on a primitive unit cell
whose basal lattice vectors are constrained to form a simple square
lattice having a specified lattice parameter $a$.
This allows us to isolate the effects of epitaxial strain
from surface and interface effects which would also be present in a real film.

\subsection{Structures: PTO and PZO}

For ferroelectric PTO and PZO we assume tetragonal P4mm symmetry
with 5 atoms per unit cell (point group C$_{4v}$).
We take the normal to the film to be the 4-fold axis along $z$.
In the paraelectric case, there is an additional mirror symmetry
perpendicular to z, resulting in point group D$_{4h}$.
The degrees of
freedom that are relaxed to obtain the lowest-energy configuration
are the magnitude $c$ of the lattice vector along $z$ and the
atomic displacements, all along z,
that are consistent with the symmetry constraints..
The calculations are carried out as a function of the epitaxial
strain, i.e., in-plane lattice constant $a$.
Restricting the calculations to a single tetragonal phase allows
for an examination of the intrinsic properties of the crystal in
a continuous manner, without considering abrupt changes due to
possible strain-induced phase transitions.

Domain-wall energies and the barriers to domain-wall
motion are studied using the methods outlined in Meyer and
Vanderbilt\cite{Meyer2002}.  For pure PTO and PZO, the domain wall
is modeled using
an $N\times1\times1$ supercell containing $N/2$ ``up'' unit cells
and $N/2$ ``down'' cells.  It is found that the domain walls are
sufficiently thin that the energies are converged for $N$=6; however,
supercells of size $N$=8 are used to investigate the local structure.
Epitaxial constraints are
applied to the domain-wall calculations, as follows.
The $c$ parameter
is held fixed at the computed value for the bulk ferroelectric
structure, appropriate for an isolated domain wall far from any
surfaces or interfaces, and the in-plane lattice parameters are
fixed at $Na$ and $a$, which define the strain state of the system.
The symmetries of the supercell are selected
such that the domain wall is centered either on a PbO plane or on
a TiO$_{2}$ (or ZrO$_2$) plane.
In the simplest scenario, the difference of the energies of
these two configurations would give the idealized barrier for domain-wall
motion.

\subsection{Structures: PZT}

Random alloys such as PZT (PbZr$_{1-x}$Ti$_{x}$O$_{3}$)
pose a challenge to first-principles methods
because the atomic arrangement is
not naturally compatible with periodic boundary
conditions.  One approach to addressing this problem is to construct
large supercells in which the cation arrangement approaches randomness,
either by randomizing the occupation of each cation site in
the supercell or by the application of techniques for constructing
so-called special quasirandom structures\cite{beck2008}.
Using effective-Hamiltonian methods or parameterized interatomic
potentials, it may be possible go to quite large supercells
\cite{Bellaiche2000-2,Grinberg2004},
but one is typically limited to rather small supercells when using
direct first-principles methods.

To study PZT at $x=0.50$, we here adopt this supercell approach,
considering two minimal supercells:
(i) a superlattice structure with alternating
layers of PTO and PZO in the (100) direction, and (ii) a rock-salt
structure (or ``3D chess board'') in which the chemical identity
alternates between all nearest B--B neighbors in the ABO$_3$
structure.  We again constrain the in-plane lattice constants to
equal those of an assumed cubic substrate, and impose orthorhombic C$_{2v}$
or tetragonal C$_{4v}$ symmetry for the (100)-- and rock-salt--ordered
superlattices respectively.  In the paraelectric case, when the polarization 
vanishes, the corresponding symmetry labels are D$_{2h}$ and D$_{4h}$.

\subsection{Computational Methods\label{CM}}

Density-functional theory calculations are performed using the
local-density approximation\cite{Kohn1965,Ceperley1980} as
implemented in the PWscf code package\cite{QE40}.
Ultrasoft pseudo\-potentials\cite{Vanderbilt1990,footnote1} are used
in place of the all-electron ion potentials.
The Kohn-Sham wave functions are expressed in terms of a plane-wave
expansion, and the Brillouin zone is sampled using the
Monkhorst-Pack algorithm\cite{Monkhorst1976}.  The coherent reversal
calculations were performed with a 35\,Ry plane-wave cutoff and
6$\times$6$\times$6 k-point sampling, while the domain-wall
calculations were carried out using a 30\,Ry cutoff and
4$\times$4$\times$2 sampling; in all cases these were chose to
ensure that the forces are converged to better than 15\,meV/\AA.
The Berry-phase technique\cite{King-Smith1993}
is used to calculate the polarization of the structures.

With these choices
the lattice parameters of known structures are calculated to be
near experimental values for PTO,
$a_{0}^{\rm exp}=3.90$\,\AA\ and
$c_{0}^{\rm exp}=4.15$\,\AA \cite{Jona1993,Shirane1956}.
In particular, we obtain equilibrium lattice parameters for the
tetragonal ferroelectric phases of PTO and PZO of
$a^{\rm PTO}_0=3.86$\,\AA, $c^{\rm PTO}_0=4.02$\,\AA,
$a^{\rm PZO}_0=4.09$\,\AA, and $c^{\rm PZO}_0=4.16$\,\AA.
The calculated lattice parameters for the PZT structures
(with symmetry imposed as explained above)
are $a^{\rm PZT}_0=3.97$\,\AA\ and $c^{\rm PZT}_0=4.09$\,\AA,
consistent with the predictions of Vegard's law.  All epitaxial
strain values reported in this paper are defined relative to the
corresponding ground-state $a_0$ values given here.

\section{Results}\label{results}

\subsection{Structural properties of paraelectric and ferroelectric
states \label{sec:strucprop}}

\begin{figure}
\begin{center}
\includegraphics[width=3.0in]{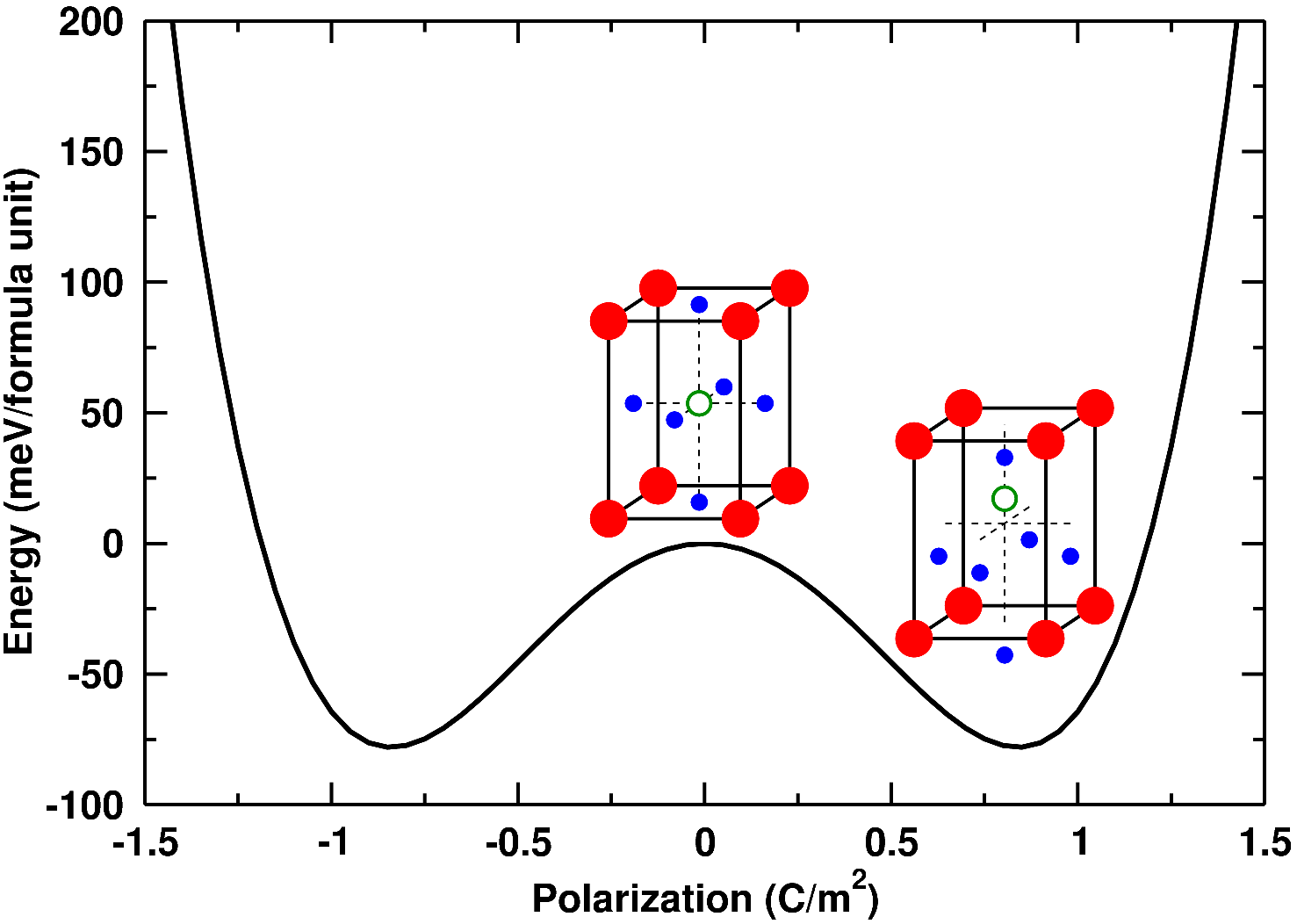}
\caption{Double-well potential of PTO at zero in-plane strain.
The ferroelectric state (minimum sketched at right) has point-group symmetry
C$_{4v}$; the paraelectric state (saddle point sketched at center)
has D$_{4h}$.}
\label{dblwell}
\end{center}
\end{figure}

Figure \ref{dblwell} shows a sketch of the double-well potential
of a typical ferroelectric, indicating the structure of one of
the local ferroelectric minima and of the paraelectric saddle-point
configuration.
We calculate the energy, $c/a$ ratio, and polarization for
these ferroelectric and paraelectric states in PTO, PZO, and PZT
by relaxing $c$ as well as all internal coordinates
subject to the appropriate symmetries specified in the previous
section.

\begin{figure}[b]
\begin{center}
\includegraphics[width=3.25in]{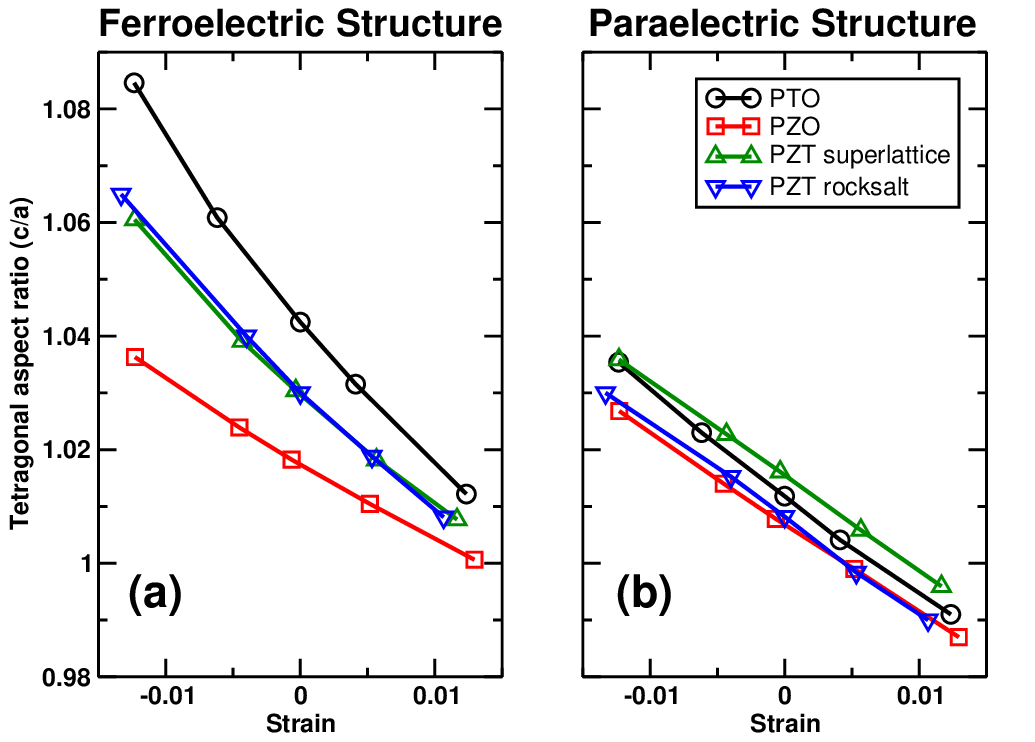}
\caption{Lattice-constant ratio $c/a$ versus applied epitaxial strain,
obtained by relaxing in the ferroelectric (a)
or paraelectric (b) state.}
\label{aspectratio}
\end{center}
\end{figure}

In Fig.~\ref{aspectratio} we present our computed values of the
$c/a$ ratio as a function of epitaxial strain for both symmetries.
Applying a compressive in-plane strain causes $c$ (and
$c/a$) to increase, which corresponds to the common situation where
the Poisson ratio is positive.
This increase is roughly linear and
composition-independent in the paraelectric case,
shown in Fig.~\ref{aspectratio}(a), where the polarization plays no role.
However, in the ferroelectric case, the curves are
steeper, more non-linear, and more composition-dependent, as shown
in Fig.~\ref{aspectratio}(b), where the electric polarization is enhanced
by compressive in-plane strain.

\begin{figure}
\begin{center}
\includegraphics[width=3.25in]{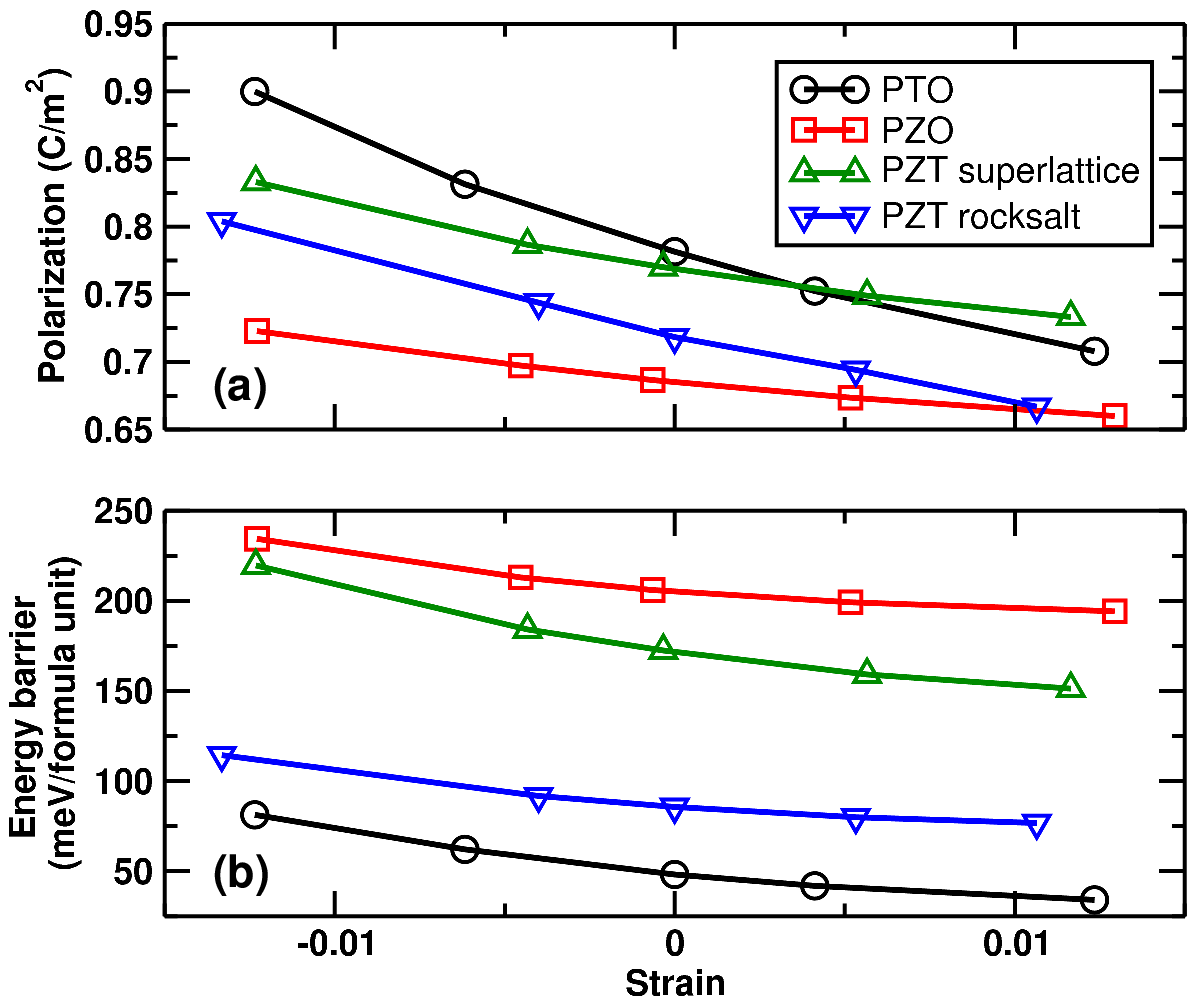}
\caption{Polarization (a) and energy-barrier height (b) versus applied
epitaxial strain in tetragonal PTO-PZO systems.}
\label{pande}
\end{center}
\end{figure}

The enhancement of polarization by compressive epitaxial strain
can be seen clearly in Fig.~\ref{pande}(a).
The polarization is generally largest
in PTO, smallest in PZO, and intermediate in PZT.  This ordering is
consistent with the variation in the $c/a$ ratios presented
in Fig.~\ref{aspectratio}.
However, Fig.~\ref{pande}(b)
shows that the height of the barrier is
actually larger for PZO than for PTO, with the PZT alloys again lying
at intermediate values.

\subsection{Double-well potential and domain reversal\label{sec:dom}}

A double well potential as a function of polarization is
obtained as a fourth order polynomial
\begin{equation}
E(P)=aP^{2}+bP^4,\label{landauisheqn}
\end{equation}
where $E$ is the energy (relative to the paraelectric saddle point)
and $P$ is the polarization.  Knowing the values $P_1$ and $E_1$
that characterize the ferroelectric minimum, Eq.~(\ref{landauisheqn})
can be rearranged to find $a=2E_{1}/P_{1}^{2}$ and $b=-E_{1}/P_{1}^{4}$.
Figure \ref{dblwell} shows the double-well potential
of ferroelectric PTO at zero in-plane strain.
The various double-well
potentials can conveniently be analyzed by fitting them to this
simplified functional form\cite{rabe2007}.
For example, the shapes of the double-well potentials are seen to be
rather different
for PTO and PZO, with the latter having a deeper minimum that however
occurs at a smaller value of polarization.

Introducing the electric enthalpy $F=E-{\cal E}P$, where
$\cal E$ is the electric field, and recalling that
${\cal E} =dE/dP$ corresponds to the slope of the $E(P)$ curve,
we obtain
\begin{equation}
{\cal E}_{\rm c}=(4/3)^{3/2}E_{1}/P_{1} \label{ecequ}
\end{equation}
for the ideal intrinsic coercive field, defined as the maximum
slope attained between the minimum and saddle point.
It should be emphasized that this is an artificial quantity that may
be orders of magnitude larger than physical coercive fields; it
corresponds to the field at which the minority domain ceases to be
defined theoretically, whereas physical coercive fields are defined by
the field at which domain walls become unpinned.  In comparison,
Lee {\it et al.}\ find coercive fields of $\sim$25\,MV/m
in PbZr$_{0.2}$Ti$_{0.8}$O$_{3}$ thin films of
100\,nm thickness.

\begin{figure}[t]
\begin{center}
\includegraphics[width=3.0in]{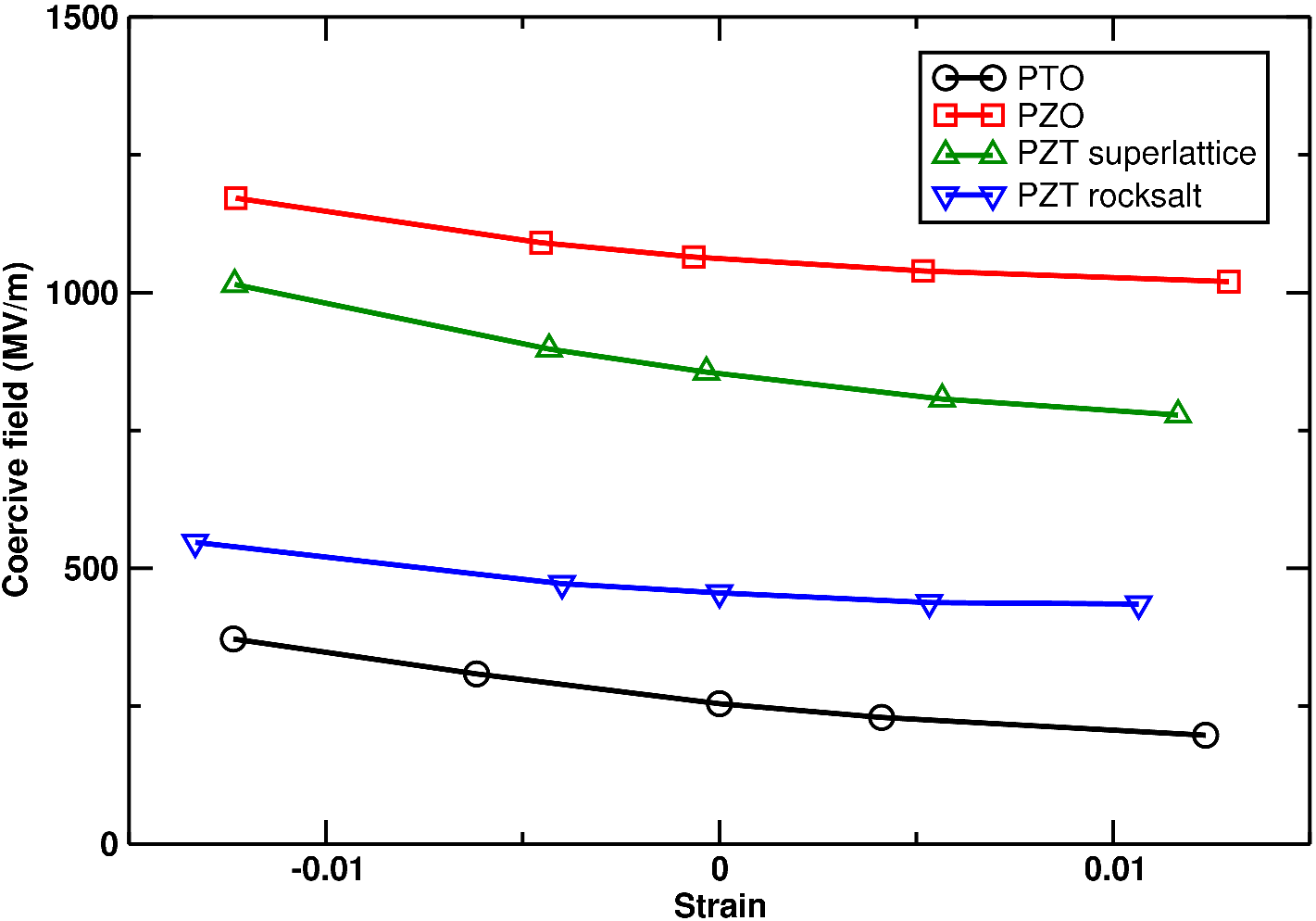}
\caption{Ideal coercive field versus epitaxial strain as determined
from Eq.~(\ref{ecequ}) using the data from Fig.~\ref{pande}.}
\label{ecoer}
\end{center}
\end{figure}

Using Eqs.~(\ref{landauisheqn}-\ref{ecequ})
and the data in Fig.~\ref{pande}, the
coercive field is calculated as a function of applied strain and
plotted in Fig.~\ref{ecoer}.  The ideal coercive field can be
seen to be much larger for PZO than for PTO (with PZT at
intermediate values).
This is consistent with the trends
identified earlier, since $E_1$ increases while $P_1$ decreases
in going from PTO to PZO, and from Eq.~(\ref{ecequ}) it is
obvious that both of these effects tend to increase the ideal
coercive field.
The observed strain dependence of the polarization
and coercive field indicate that the application of compressive
strain expands the hysteresis loops and applying a tensile
strain contracts them.

\begin{figure}[b]
\begin{center}
\includegraphics[width=3.25in]{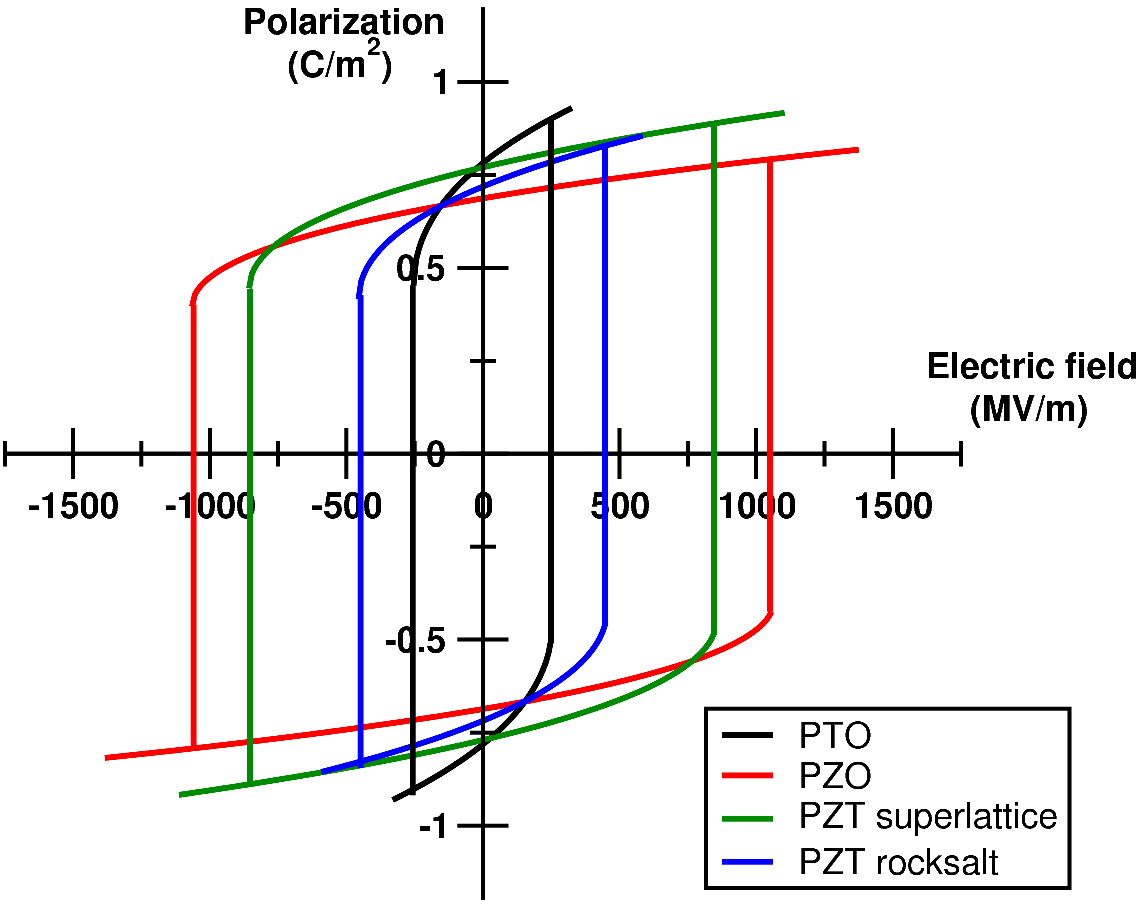}
\caption{Hysteresis curves calculated from Eq.~(\ref{landauisheqn})
at zero epitaxial strain.}
\label{hyst}
\end{center}
\end{figure}

From the simple model of Eq.~(\ref{landauisheqn}) it is also
straightforward to plot the polarization as a function of
electric field in order to obtain ideal hysteresis curves such as those
plotted in Fig.~\ref{hyst}.
Again we emphasize that this corresponds
to coherent domain reversal, i.e., a scenario in which the reversal
happens simultaneously everywhere without loss of crystal
periodicity.  The ideal coercive field can be seen
to be much larger for PZO than for PTO, consistent with Fig.~\ref{ecoer}.

\subsection{Domain walls \label{sec:dw}}

\begin{figure}
\begin{center}
\includegraphics[width=3.25in]{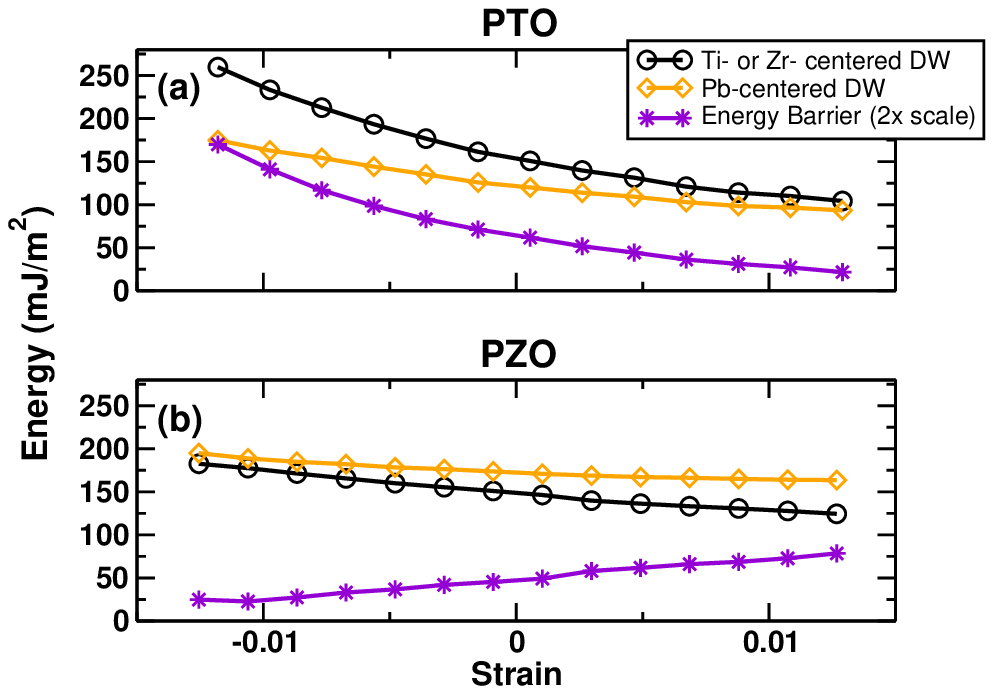}
\caption{Domain wall energy and barrier to domain-wall motion in
strained PTO (a) and PZO (b) films. The energy barrier is scaled $2\times$.}
\label{PTPZdw}
\end{center}
\end{figure}

The energies of domain walls are calculated for PTO, PZO, and the
(100) superlattice PZT.  For all three crystals the supercell
belongs to the D$_{2h}$ point group.
We constrain the domain-wall location by enforcing an inversion
symmetry about a center in the specified plane, and relax all
other internal degrees of freedom.  For PTO and PZO the domain-wall
energies are shown on the right axis of Fig.\ \ref{PTPZdw}.
These increase with compressive strain, for which
the polarization is larger and therefore the cost of introducing
an unpolarized layer is larger.
A simple estimate of the energy barrier to domain-wall motion is
the difference between the two domain-wall energies, which is
plotted along the left axis of Fig.\ \ref{PTPZdw}.
These barriers have a strain dependence quite different
in PTO than in PZO.  We shall discuss this behavior further in
Sec.~\ref{sec:dwprop}.

\begin{figure}
\begin{center}
\includegraphics[width=3.25in]{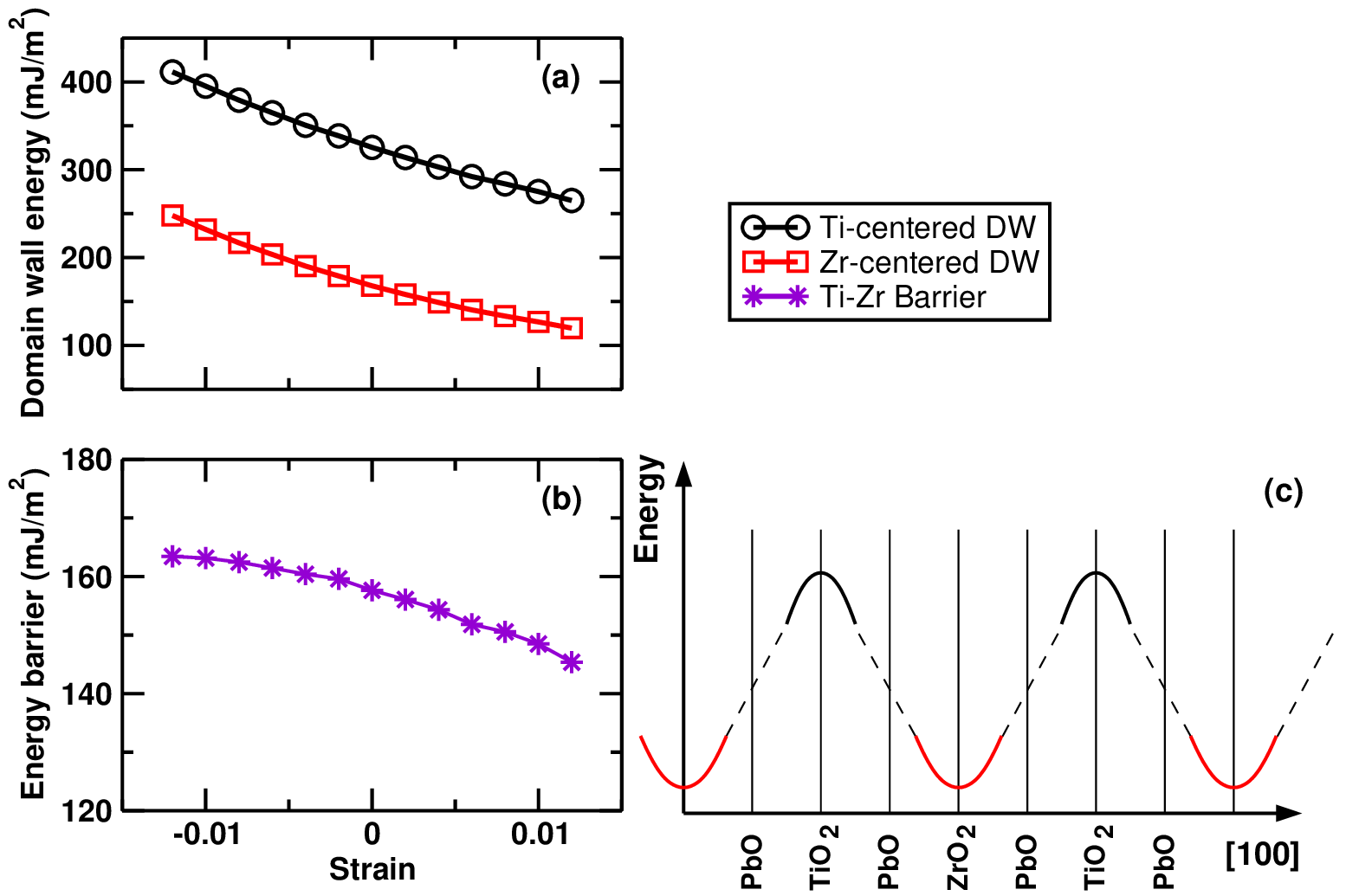}
\caption{Domain-wall energy (a) and barrier to domain-wall motion (b) in
a strained (100) PZT superlattice.  Sketch of energy landscape
emerging from the calculations (c).
}
\label{PZTdw}
\end{center}
\end{figure}

The domain-wall energies are shown for the (100) PZT superlattice
in the top frame of Fig.\ \ref{PZTdw}(a).  Here it is possible to
specify a domain wall located on a ZrO$_{2}$ or TiO$_{2}$ plane,
but not on a PbO plane, because the compositional order makes it
impossible to choose an inversion center in a PbO plane.  As is
evident in the figure, the Zr-centered domain wall is lower in
energy at all strains.  If we choose initial atomic
coordinates corresponding to a domain wall that is positioned
on a PbO plane, we find that the atomic relaxations spontaneously
carry the domain wall toward the nearest ZrO$_{2}$ plane,
consistent with the fact that it is not pinned by symmetry
constraints.  Figure \ref{PZTdw}(c) shows a sketch of the resulting
energy landscape for domain-wall motion in this superlattice system.
We have not attempted to calculate (or even to precisely define) the
energy surface between the TiO$_{2}$ and ZrO$_{2}$ planes, so the
dashed lines in the Fig.\ \ref{PZTdw}(c) are just conjectural.

\section{Discussion \label{D}}

In this section we take a closer look at the results presented in
Sec.~\ref{results}, seeking to explain some of those results in
terms of interactions at the atomistic level.  The influence of local
strain on the structure will be examined.

\subsection{Polarization and polarization reversal}

\subsubsection{Dependence of polarization on composition}

In Fig.\ \ref{aspectratio}(a) it can be seen that the $c/a$
ratio for ferroelectric PZT obeys Vegard's law quite well,
being very close to the average of PTO and PZO at all strains.
However, this is not the case for the polarization in
Fig.~\ref{pande}(a), where the PZT superlattice and rock-salt
structures lie above and below the average, respectively.

This can be understood by inspecting the local strains and how
they impact the properties of the crystal.  Looking
first at the superlattice case, while the periodicity is
$2a^{\rm PZT}_0=7.94$\,\AA~in the (100) direction, the Pb atoms
are not uniformly spaced.  The spacing is 3.81 and 4.14\,\AA\ across
the TiO$_{2}$ and ZrO$_{2}$ planes respectively.
Defining the local strain in the
(100) direction relative to the equilibrium lattice parameters
($a^{\rm PTO}_0=3.86$ and $a^{\rm PZO}_0=4.09$\,\AA), the PTO layer
has a strain of $-$1.3\% while that of the PZO layer is +1.2\%.
Using the data for strained PTO and PZO in Fig.~\ref{pande}(a) and
applying these local strains, we can estimate the polarization
contribution to be 0.91 and 0.66\,C/m$^{2}$ from the
PTO and PZO layers respectively.  The average of these is
0.79\,C/m$^{2}$, which is very close to the calculated Berry-phase value
of 0.77\,C/m$^{2}$.
A similar analysis applies to the rock-salt case.
The Pb separation in both the (100) and (010)
directions is 3.97\,\AA, corresponding to a strain of $\pm$2.8\%
in the PTO and PZO cells respectively.  Extrapolating
linearly using the data in Fig.~\ref{pande}(a), we expect
contributions from the PZO and PTO components to be 0.62 and
0.77\,C/m$^{2}$.  This averages to 0.70\,C/m$^{2}$, which is
again very close to the calculated value of 0.72\,C/m$^{2}$.

\subsubsection{Local distortions and partial distribution functions}

\begin{figure}
\begin{center}
\includegraphics[width=3.25in]{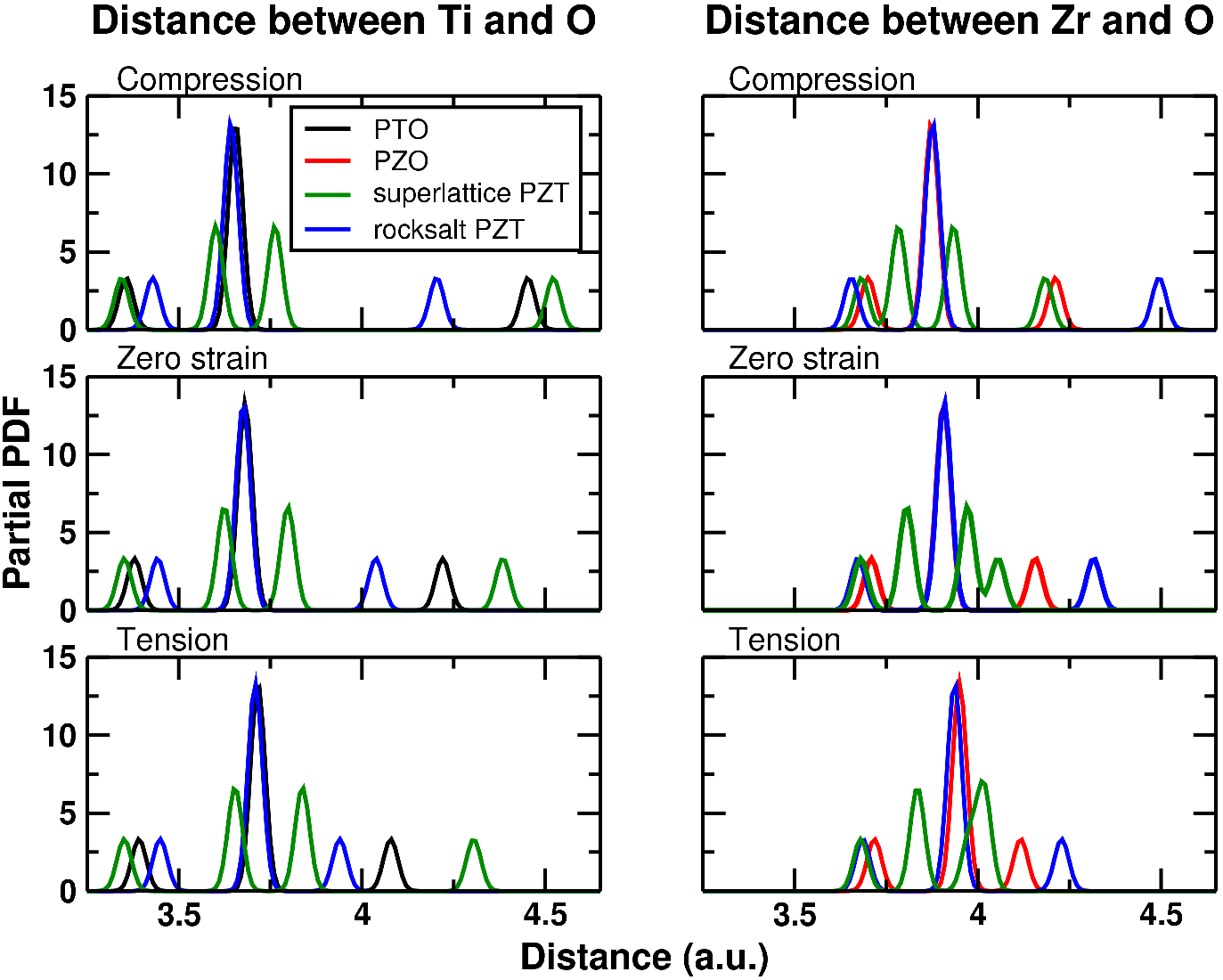}
\caption{Partial PDFs for Ti--O (left) and Zr--O (right) distances.
Top, middle, and lower panels show applied epitaxial strains of
$\epsilon=-1.2$\% (compressive), 0, and +1.2\% (tensile), respectively.}
\label{BOpdf}
\end{center}
\end{figure}

The local strain arguments above suggest that
in the superlattice the TiO$_{2}$ layer contributes more
to the net polarization than the ZrO$_{2}$ layer, whereas in
the rock-salt structure the Ti and Zr cells make approximately
equal contributions.  The contributions of the Ti and Zr ions
are roughly proportional to their displacements relative to the
surrounding oxygen octahedra.\cite{footnote2,Zhong1994}
The calculated relative displacements are presented as partial
pair distribution functions (PDFs)\cite{footnote3}
for Ti--O and Zr--O neighbors in Fig.~\ref{BOpdf}.  The first peak
at the shortest distance in each of the distributions corresponds to
the Ti or Zr bonded to the O atom located in the
nearest $xy$-oriented (basal) PbO plane.
This Zr--O bond length is almost identical in the two PZT structures,
while the Ti--O bond length is about 2.5\% shorter in the superlattice than
the rock-salt structure.  This supports the hypothesis that in
the superlattice structure the TiO$_{2}$ layer contributes more
to the net polarization than does the ZrO$_{2}$ layer.  In addition,
the Ti--O bond is shorter in the
superlattice and the spacing between PbO layers is smaller,
leading to a larger energy barrier in the superlattice as
compared to the rock-salt structure.

The sensitivity of a crystal's polarization to its strain can also be
related to the atomic structure.  From Fig.~\ref{pande} it is
clear that PTO is more sensitive to applied strain than is the
case for PZT or PZO.  The PDFs indicate that the change in Ti--O
bond length under applied load in PTO is larger than in PZT, and larger
than that of Zr--O bonds in PZT or PZO.
It is this large change in bond length
that leads to the change in PTO polarization with applied strain.

\begin{figure}
\begin{center}
\includegraphics[width=3.25in]{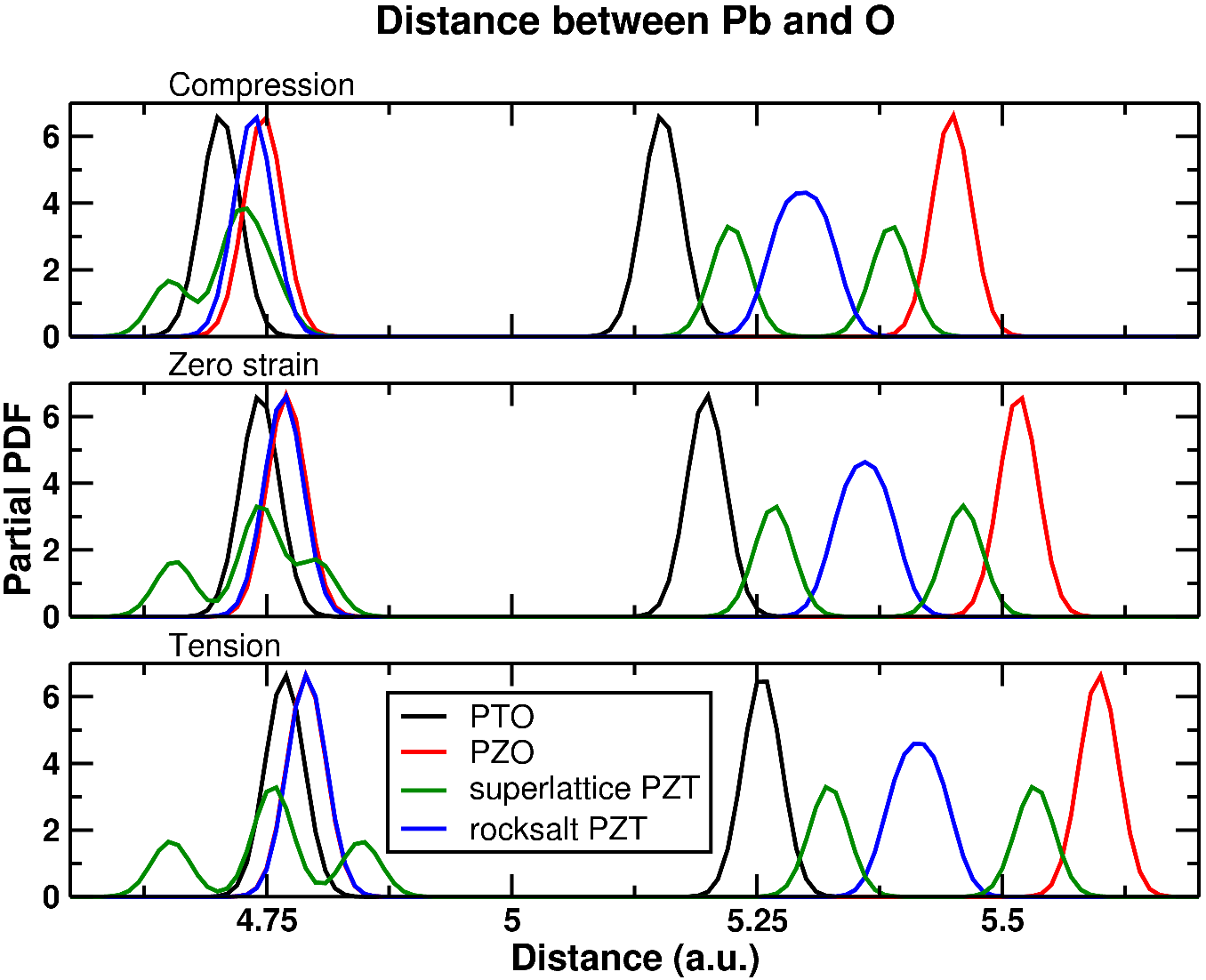}
\caption{Partial PDF of Pb--O distances.
Top, middle, and lower panels show applied epitaxial strains of
$\epsilon=-1.2$\%, 0, and +1.2\%, respectively.}
\label{AOpdf}
\end{center}
\end{figure}

A partial PDF of Pb--O separations is shown
Fig.~\ref{AOpdf}. The shorter distances (those below 5\,a.u.)
correspond to Pb and O atoms lying in the same $xz$ or $yz$ plane.
The superlattice has enough flexibility that the shortest of these
can remain almost independent of strain, while the others
show a modest shift in bond length with strain.
On the other hand, the longer bonds in the right part of the figure,
which correspond to Pb--O bonds in the basal plane, show much
larger shifts because they have little choice but to scale
linearly with the $a$ lattice parameter.  In rock-salt PZT
the structure has so little flexibility that its Pb--O bond
distances are essentially just averages of the PTO and PZO ones,
as can be seen in Fig.~\ref{AOpdf}.  In the PZT superlattice,
on the other hand, the peak splits strongly because there are
two PbO interplanar distances, with the peaks near 5.45 and
5.25\,a.u.\ corresponding to Pb bonds with O atoms in a
TiO$_2$ or ZrO$_2$ plane, respectively.
Understanding the spatial relationships between the Pb and the
basal O atoms helps to further illuminate the nature of the
partial PDFs for Ti--O and Zr--O shown in Fig.~\ref{BOpdf}.

\begin{figure}
\begin{center}
\includegraphics[width=3.25in]{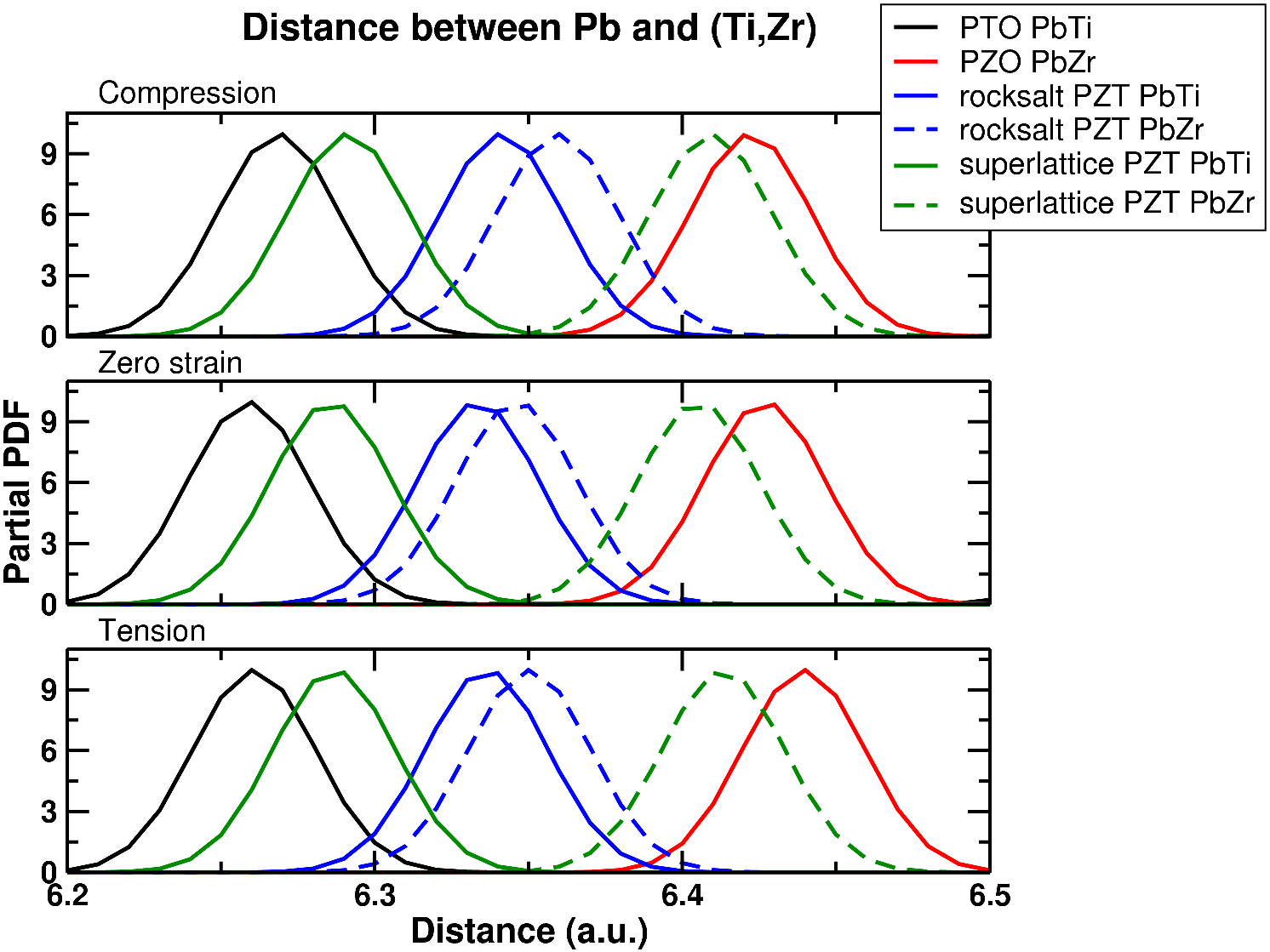}
\caption{Partial PDF of Pb--Ti and Pb--Zr distances.
Top, middle, and lower panels show applied epitaxial strains of
$\epsilon=-1.2$\%, 0, and +1.2\%, respectively.}
\label{ABpdf}
\end{center}
\end{figure}

The partial PDFs in Fig.~\ref{ABpdf} show the Ti--Pb and Zr--Pb
separations.
The distributions look very reminiscent of those shown for
Pb--O distances in the right part of Fig.~\ref{AOpdf}, except that
here there is almost no shift with strain.
The peak associated with the
rock-salt structure splits because the ferroelectric distortions of the
Ti and Zr atoms differ, but the splitting is quite small
($<$ 0.01 a.u.).  The Pb and Ti or Zr atoms are not bonded to one
another, but interact indirectly through shared O atoms.

\subsection{Domain-wall properties\label{sec:dwprop}}

\begin{figure}[b]
\begin{center}
\includegraphics[width=3.0in]{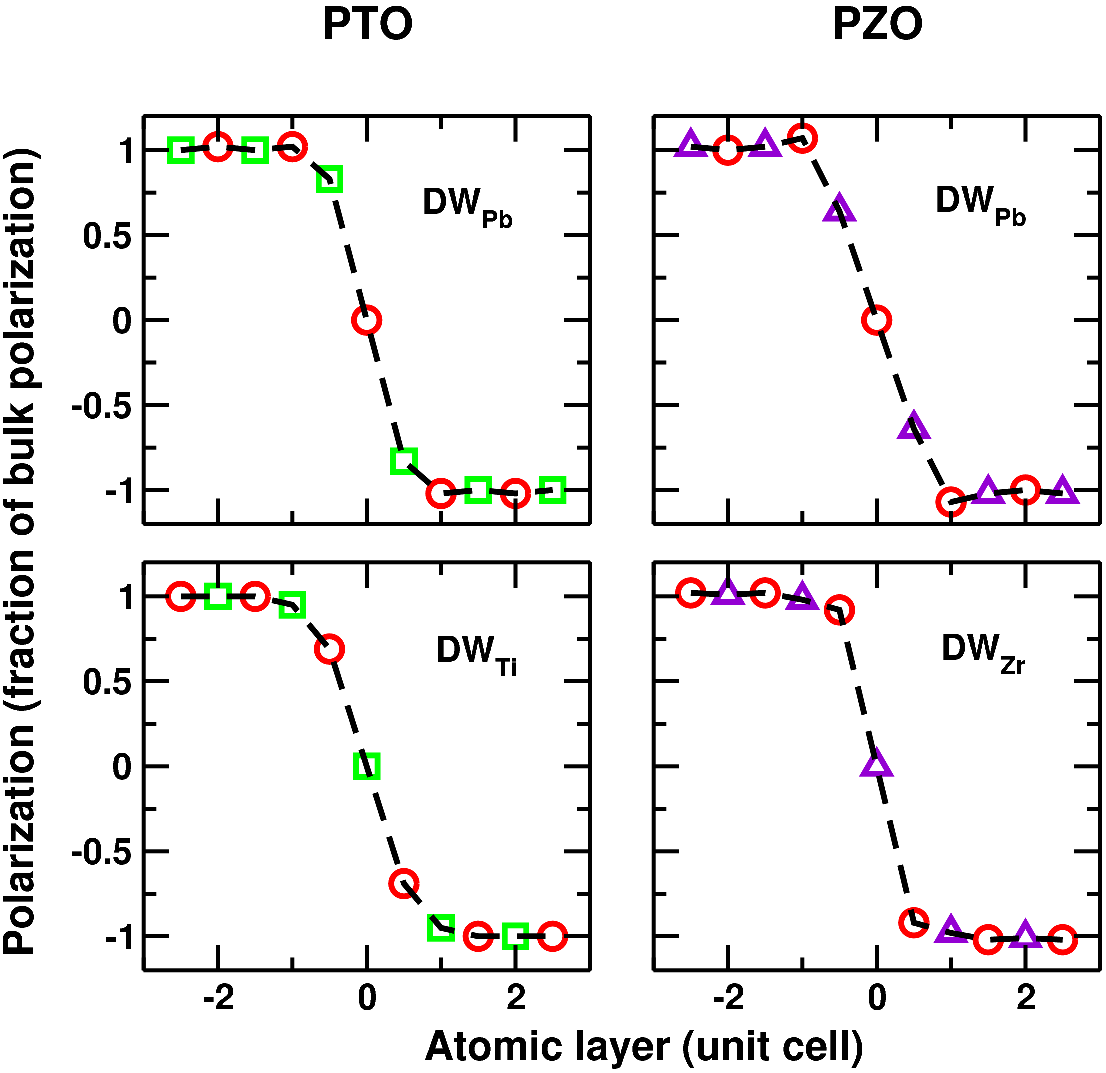}
\caption{Decomposition of polarization on a plane-by-plane
basis across domain wall.
Left, PTO; right, PZO.
Top, Pb-centered walls; bottom, Ti- or Zr-centered walls.
Symbols denote PbO (circles), TiO$_2$ (squares), and
ZrO$_{2}$ (triangles) planes.}
\label{PDW}
\end{center}
\end{figure}

\begin{figure}[b]
\begin{center}
\includegraphics[width=3.0in]{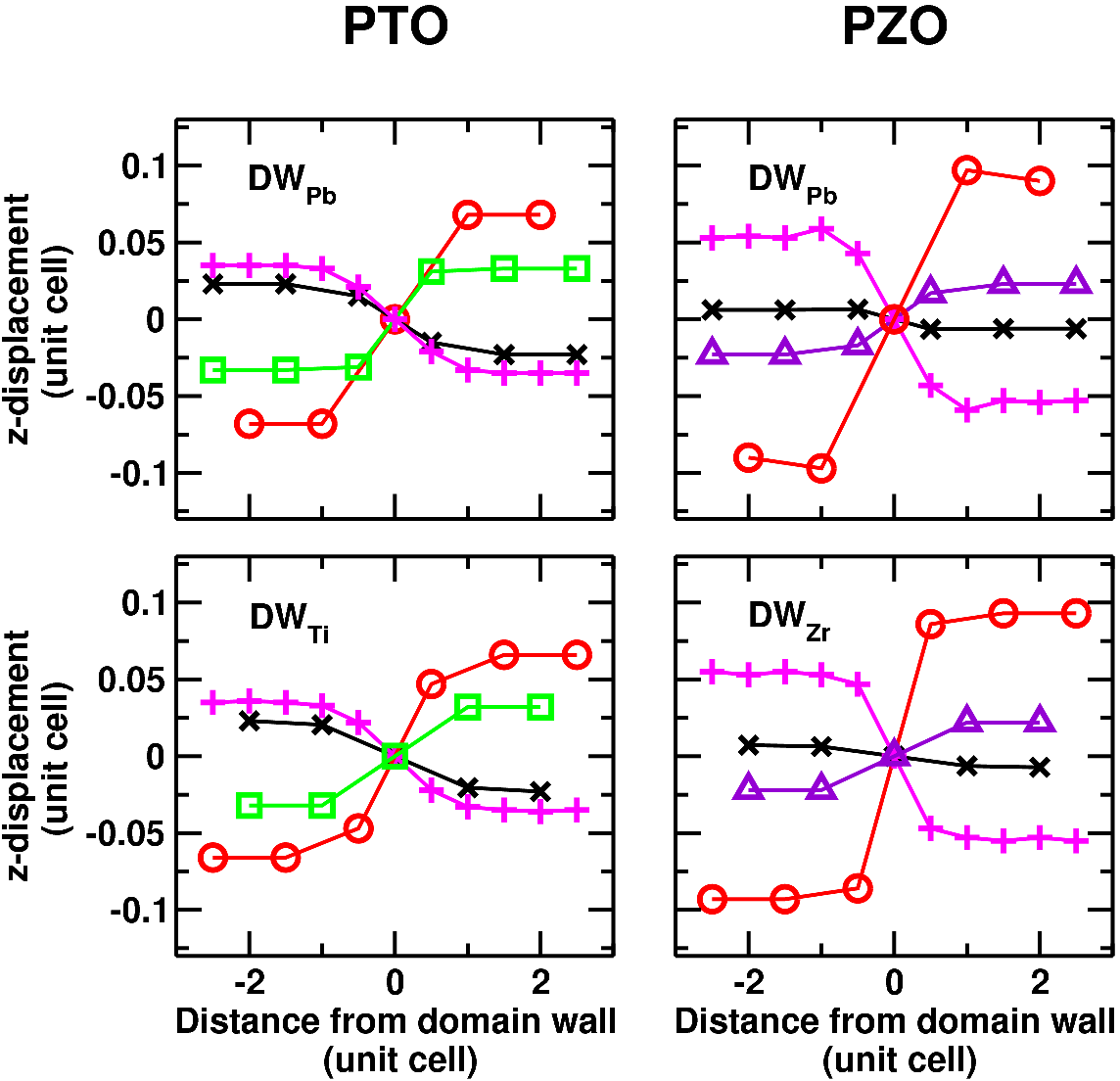}
\caption{Atomic displacements across domain walls.
Left, PTO; right, PZO.
Top, Pb-centered walls; bottom, Ti- or Zr-centered walls.
Symbols denote Pb (circles), Ti (squares), Zr (triangles),
O in the $xz$ or $yz$ plane (pluses), and O in the $xy$ plane
(crosses).}
\label{atomDW}
\end{center}
\end{figure}

We turn our attention now to the results on domain walls presented
in Sec.~\ref{sec:dw}.  We have inspected the atomistic structure
of the domain walls, and find that in all cases they are only a
few atomic planes thick.  To illustrate this, the polarization
profiles atomic displacements and obtained as one scans across
the domain wall are plotted in Figs.~\ref{PDW}-\ref{atomDW}
for both PTO and PZO at zero epitaxial strain.  The polarizations
in Fig.~\ref{PDW} are determined by the procedure outlined
in Ref.~~\onlinecite{Meyer2002} using the atomic displacements in
Fig.~\ref{atomDW} and are normalized such that the bulk polarization has
a value of 1.  Similar domain-wall thicknesses
are observed for PZT.  These results confirm those obtained for PTO
in Ref.~\onlinecite{Meyer2002}, and it is clear that the 180$^{\circ}$
domain walls are atomistically sharp, at least at zero temperature.
Since the domain walls are all about equally narrow, it seems likely
that their energies should correlate mainly with the energy cost
of introducing an atomic plane in which the ferroelectric distortions
are absent.  This hypothesis is consistent with the observed increase
of domain-wall energy with compressive strain, which makes the
bulk more ferroelectric, thus introducing a larger penalty for
a layer that is not ferroelectric.

However, the barrier to
domain-wall motion does not depend on the absolute energies of
the domain walls, but on their relative energies, as sketched in
Fig.~\ref{PZTdw}(c).
Consider first the energy barriers in PTO and PZO,
plotted in Fig.\ \ref{PTPZdw}.
The energy barrier in PTO increases substantially  moving
from tensile to compressive strain.  This is because the domain wall
centered on the TiO$_{2}$ layer is highly sensitive to strain compared to the
PbO layer.  Moving from $\epsilon=+1.2$\% to $-1.2$\% the energy of
the TiO$_{2}$ centered wall increases by 155 mJ/m$^{2}$.  This is almost
twice the increase of the PbO centered wall, which increases 81 mJ/m$^{2}$.
The barrier to domain-wall motion is related to the difference in the
domain-wall energies, and the rate of change of barrier height with
applied strain is related to the difference in the rate of change of domain
wall energies.

In comparison, consider the energy barrier in PZO.
Under compression the barrier to domain-wall motion is small because the
domain walls centered on the ZrO$_{2}$  and PbO planes have around the
same energy.
The barrier increases with applied tensile strain because the domain
wall energies at the two planes diverge from one another in the
tensile regime.  Although the domain-wall energy normally
increases with applied compressive load, the barrier to wall
motion can decrease under some circumstances, because the barrier
is a relative measure.

\begin{figure}
\begin{center}
\includegraphics[width=3.25in]{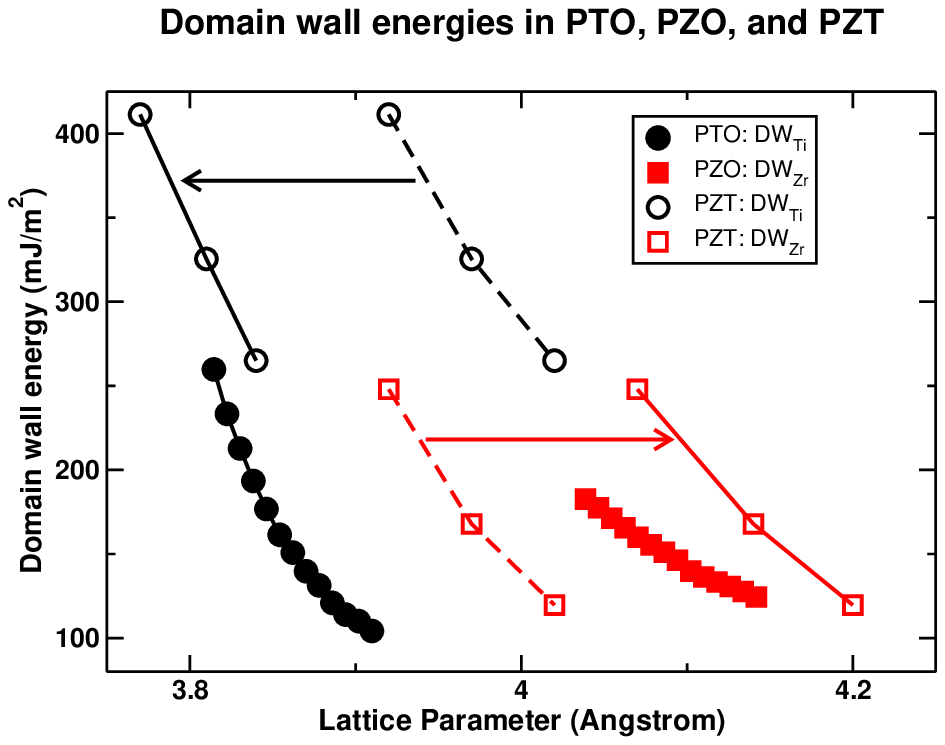}
\caption{Dependence of domain-wall energies on in-plane lattice
constant.  Solid symbols: PTO and PZO.  Open symbols: PZT, plotted
versus average lattice constant (connected by dashed lines)
or versus local lattice constant (connected by solid lines).}
\label{DWalat}
\end{center}
\end{figure}

The sensitivity of the TiO$_{2}$ layer to strain also appears
to be responsible for the large
barrier in the PZT superlattice.  This is observed by replotting
Figs.~\ref{PTPZdw} and \ref{PZTdw} in terms of the lattice
parameter, as shown in Fig.~\ref{DWalat}.  When the PZT
domain-wall energies are plotted versus the {\it average} lattice parameter
(solid symbols and open symbols connected by dashed lines),
no obvious relationship emerges.
On the other hand, the data clusters well when the domain-wall
energies are plotted versus the {\it local} lattice constant
(solid symbols and open symbols connected by solid lines),
defined in terms of the local
lattice spacing of the PbO planes in the (100) direction.  We can
thus understand that the small spacing of the TiO$_{2}$ planes
is responsible for the large domain-wall energy and the large energy
barrier for domain-wall motion.

\section{Conclusions \label{C}}

Working with the prototypical PTO/PZO system, we have shown that
the application of compressive epitaxial strain to a tetragonal
ferroelectric perovskite increases its $c/a$ aspect ratio, polarization,
barrier to coherent polarization reversal, ideal coercive field,
and domain-wall energies.  These relationships have been quantified
and related to the enhancement of the
atomic ferroelectric distortions and the strengthened Ti--O bonding.
The bond between the Ti atom and the basal O is highly sensitive
to applied strain, as observed in the calculated partial PDFs.
This sensitivity is manifested in the ordered phases of PZT,
enhancing both the polarization and barrier to domain-wall motion.

In alloys, issues of local composition and strain have been shown
to play an important role.
Although some bulk properties, such as the average lattice parameter,
obey Vegard's law, many others do not.  In PZT the
bulk polarization can be understood as an average of polarization
contributions from each 5-atom cell, in which this contribution
depends crucially both on its identity (Ti or Zr) and on the
local strain.  Similarly, we have shown that the
energy of domain walls can also be understood in terms of such
local effects, specifically the composition of the
plane on which the domain wall is centered, and the local in-plane
lattice constant for cells surrounding that plane.  These results
demonstrate how apparently complex properties of ferroelectric alloys
can be understood as averages over single-cell properties under
appropriate conditions of strain and composition.  

\section{Acknowledgments}

This work was supported by ONR Grant N00014-05-1-0054 and by the
Intel Corporation via SRC Grant 2007-VJ-1670.
Computational resources were provided in part by the National
Energy Research Scientific Computing Center (NERSC).

\bibliography{beckman_prb_2008}% Produces the bibliography via BibTeX.

\begin{thebibliography}{29}
\bibitem[{foo({\natexlab{b}})}]{footnote2}
\bibinfo{note}{The polarization is proportional to the displacement multiplied
  by the Born effective charge. The Born effective charge for Z$^{*}$(Zr) in
  PZO is around 6 and Z$^{*}$(Ti) in PTO is around 7, which is not a
  substantial difference\cite{Zhong1994}.}
\bibitem[{foo({\natexlab{c}})}]{footnote3}
\bibinfo{note}{The partial PDFs are a histograms of atomic separations, plotted
  with a Gaussian smearing having a variance of 0.02 a.u.}
\end{thebibliography}

\end{document}